\documentclass[aps,prl,twocolumn,floatfix,superscriptaddress,showpacs]{revtex4-2}
\usepackage{natbib}
\usepackage[final]{graphicx}
\usepackage{amsfonts, amsmath, amsthm, amssymb, mathrsfs} 
\usepackage{mathtools,enumerate}
\usepackage[dvipsnames, table, xcdraw]{xcolor}
\usepackage{wrapfig, tikz,url,lipsum,float}
\usepackage{dcolumn}
\usepackage{bm}
\usepackage{xspace}

\usepackage{hyperref}
\usepackage[normalem]{ulem}   

\definecolor{RoyalBlue}{HTML}{007399}
\definecolor{Seagreen}{HTML}{009999}
\definecolor{crimsonred}{HTML}{990000}

\newcommand{\braket}[1]{\left\langle #1 \right\rangle}

\newcommand{\paren}[1]{\left( #1 \right)}

\DeclareMathAlphabet\mathbfcal{OMS}{cmsy}{b}{n}

\def \ua  {u^{\alpha}}
\def \ub  {u^{\beta}}
\def \mE  {\mathcal{E}}
\def \mEK {\mathcal{E}_{\rm K}}
\def \ab {\alpha\beta}
\def \dela {\partial^{\alpha}}
\def \delb {\partial^{\beta}}
\def \grad {{\nabla}}

\def \curl {{\nabla} \times}

\def \delt {\partial_t}

\def \d{{\rm d}}

\def \Enot {E_{\rm 0}}

\def \kk  {\bm{k}}
\def \eeo {\bm{e}_{\rm 1}}
\def \eet {\bm{e}_{\rm 2}}
\def \eo {e_{\rm 1}}
\def \et {e_{\rm 2}}
\def \zh  {\hat{z}}

\def\bu {\bm{u}}
\def\uu {\bm{u}}

\def\bn {\bm{n}}
\def\bg {\bm{g}}
\def\buh {\hat{\bm{u}}}
\def\uh {\hat{{u}}}

\def\bnh {\hat{\bn}}
\def \Fg   {\bm{F}^{\rm g}}
\def \Fst   {\bm{F}^{\rm \sigma}}
\def \bx {\bm{x}}

\def\bz {\bm {z}}
\def\bzh {\hat{\bz}}

\def\bk {\bm{ k}}
\def\bq {\bm{ q}}
\def\bg {\bm{ g}}

\def\calL {\mathcal{L}}

\def \ro{\rho_{\rm 1}}
\def \rt{\rho_{\rm 2}}

\def\ra{\rho_{\rm a}}
\def\ca{c_{\rm a}}

\def \Rey  {\mbox{Re}_{\lambda}}
\def \ReL  {\mbox{Re}_{\calL}}
\def \Ga  {\mbox{Ga}}
\def \At  {\mbox{At}}
\def \Bo  {\mbox{Bo}}

\def \urms  {u_{\rm rms}}

\def \epg  {\epsilon^{\rm g}}

\def \PiK   {\Pi_{ K}}
\def \PK   {\mathscr{P}_{ K}}
\def \DK   {\mathscr{D}_{ K}}
\def \FsK   {\mathscr{F}^\sigma_{K}}

\def \FgK   {\mathscr{F}^{\rm g}_{K}}

\newcommand{\til}[1]{\widetilde{#1}_{K}}
\newcommand{\fil}[1]{{#1}_{K}}

\def \GK    {G_{\rm K}}
\def \tauL {\tau_{L}}

\def \Nb  {N_{\rm b}}
\def \kd  {k_{\rm d}}

\newcommand{\fig}[1]{Fig.~(\ref{#1})}
\newcommand{\subfig}[2]{Fig.~(\ref{#1}#2)}

\newcommand{\Eq}[1]{Eq.~(\ref{#1})}

\newcommand{\tabref}[1]{Table.~(\ref{#1})}

\newcommand{\REM}[1]{{{}}}

\newcommand{\ftinjrate}{In the Kolmogorov theory of single phase turbulence, 
energy dissipation rate per unit mass is used.  Since the problem at hand 
is multiphase flow, we use injection rate per unit volume.
They are simply related to each other by the factor of mean
density $\ra$.}

\begin{document}

\title{Kolmogorov Turbulence Coexists with Pseudo-Turbulence \\ in Buoyancy-Driven Bubbly Flows} 
\author{Vikash Pandey}
\affiliation{Tata Institute of Fundamental Research,  Gopanpally, Hyderabad 500046}
\affiliation{Nordita, KTH Royal Institute of Technology and
Stockholm University, Hannes Alfv{\'e}ns v\"ag 12, 10691 Stockholm, Sweden}
\author{Dhrubaditya Mitra}
\affiliation{Nordita, KTH Royal Institute of Technology and
Stockholm University, Hannes Alfv{\'e}ns v\"ag 12, 10691 Stockholm, Sweden}
\author{Prasad Perlekar} 
\affiliation{Tata Institute of Fundamental Research,  Gopanpally, Hyderabad 500046}
\begin{abstract}
We investigate the spectral properties of buoyancy-driven bubbly flows. Using 
high-resolution numerical simulations and phenomenology of homogeneous 
turbulence,  we identify the relevant energy transfer mechanisms. 
We find: (a)  At a high enough Galilei number (ratio of the buoyancy to viscous forces) the 
velocity power spectrum shows the Kolmogorov scaling with a power-law exponent $-5/3$ 
for the range of scales between the bubble diameter and the dissipation scale ($\eta$). 
(b) For scales smaller than  $\eta$, the physics of pseudo-turbulence is recovered.
\end{abstract}
\maketitle
The flow behind an array of cylinders or a grid, either moving or stationary, provides an
ideal testbed to verify and scrutinize the statistical theories of 
turbulence~\cite{batchelor_1953}.
What is the flow generated when a fluid is stirred by a dilute suspension of bubbles
as they rise due to buoyancy?
This question has intrigued researchers for the past three decades due to their occurrence in both 
industrial and natural processes~\cite{magnaudet_2000,mudde_rev_2005,cecc10,
riss18,said_rev_2019,mathai_rev_2020}. 
Experiments \cite{lance_1991,martinez_2010,risso_legendre_2010,Ma_2021,
vivek_2016,almeras2017} and numerical simulations \cite{pandey_2020,pandey_22,chibbaro_2021}
show that flows generated by dilute bubble suspensions 
are chaotic and originate due to the interplay of viscous, inertial, and
surface tension forces.
The complex spatio-temporal flow is called ``pseudo-turbulence" or ``bubble induced
agitation"~\cite{mudde_rev_2005,riss18}.

As is typical for chaotic flows, pseudo-turbulence is characterized by
the power spectrum of its velocity fluctuations
$E(k)$, which shows a power law scaling $E(k) \sim k^{-\alpha}$ 
with an exponent  $\alpha \gtrsim 3$ in the wavenumber range $k \gtrsim \kd$ where
$\kd = 2\pi/d$ and $d$ is the bubble
diameter~\cite{lance_1991,vivek_2016}.
Lance \& Bataille~\cite{lance_1991} argued that the balance of energy production with 
viscous dissipation may explain the observed scaling.
Recent numerical studies conducted for experimentally  accessible Galilei numbers 
$\Ga$  (the ratio of buoyancy to viscous dissipation), show that the net production has 
contributions both from the advective nonlinearity and the surface
tension~\cite{pandey_2020,pandey_22,chibbaro_2021,Ma_2021,pandey_22}.

In homogeneous and isotropic turbulence (HIT) the energy injected at an integral
scale $\calL$ is transferred to dissipation scale $\eta \ll \calL$,  via
the advective interactions without dissipation
while maintaining a constant energy flux.
This intermediate range of scales between $\eta$ and $\calL$  is 
called the inertial range.  
At scale smaller than $\eta$ 
the advective interactions balance viscous dissipation~\cite{frisch}.
Clearly within the phenomenology of homogeneous and isotropic turbulence, 
pseudo-turbulence is a dissipation range phenomena with the additional
complexity due to surface tension forces.
Is it possible to have an inertial range  in buoyancy
driven bubbly flows?

In this paper, we present state-of-the-art direct numerical simulations of buoyancy
driven bubbly flows, at high resolution, which allows us to access $\Ga>1000$ which
has never been achieved before in either experiments or numerical simulations.
Our multiphase simulations model a dilute suspension of 
``gas" bubbles of lighter phase 
(density $\ro$)  dispersed in the heavier  ``liquid'' phase (density $\rt$).
The density contrast is parametrised by the Atwood number,
  $\At\equiv (\rt-\ro)/(\rt+\ro)$.
  We consider both small ($0.04$) and large ($0.8,0.98$) values for $\At$. 
  We use two different codes for these two cases. 
  In both of these cases, we find, for the first time, a direct evidence for
  Kolmogorov scaling, $E(k) \sim k^{-5/3}$, for
  $ \kd \leq  k \lesssim 1/\eta $.  
For scales smaller than $\eta$, the physics of
pseudo-turbulence is recovered. 
By analyzing the scale-by-scale energy budget we uncover the
mechanism by which the Kolmogorov scaling emerges:
for high enough $\Ga$, for both small and larger $\At$,
there is an intermediate range of scales over
which the contribution from advection dominates
over all other contributions (including surface tension) in the kinetic energy budget.
This is the range over which  Kolmogorov scaling is observed. 
\begin{figure*}[!htb]
  \includegraphics[width=0.24\linewidth]{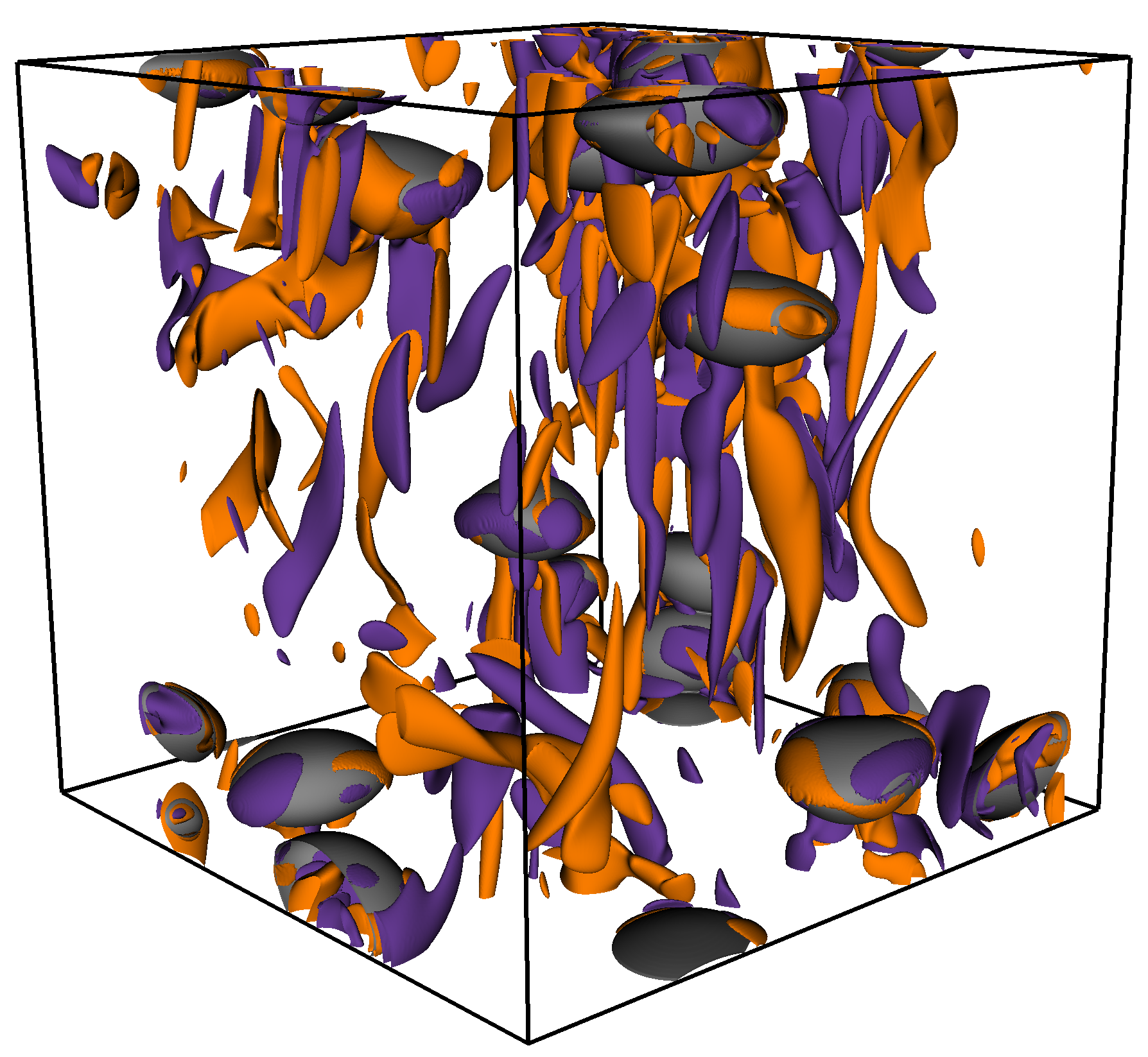}
  \includegraphics[width=0.24\linewidth]{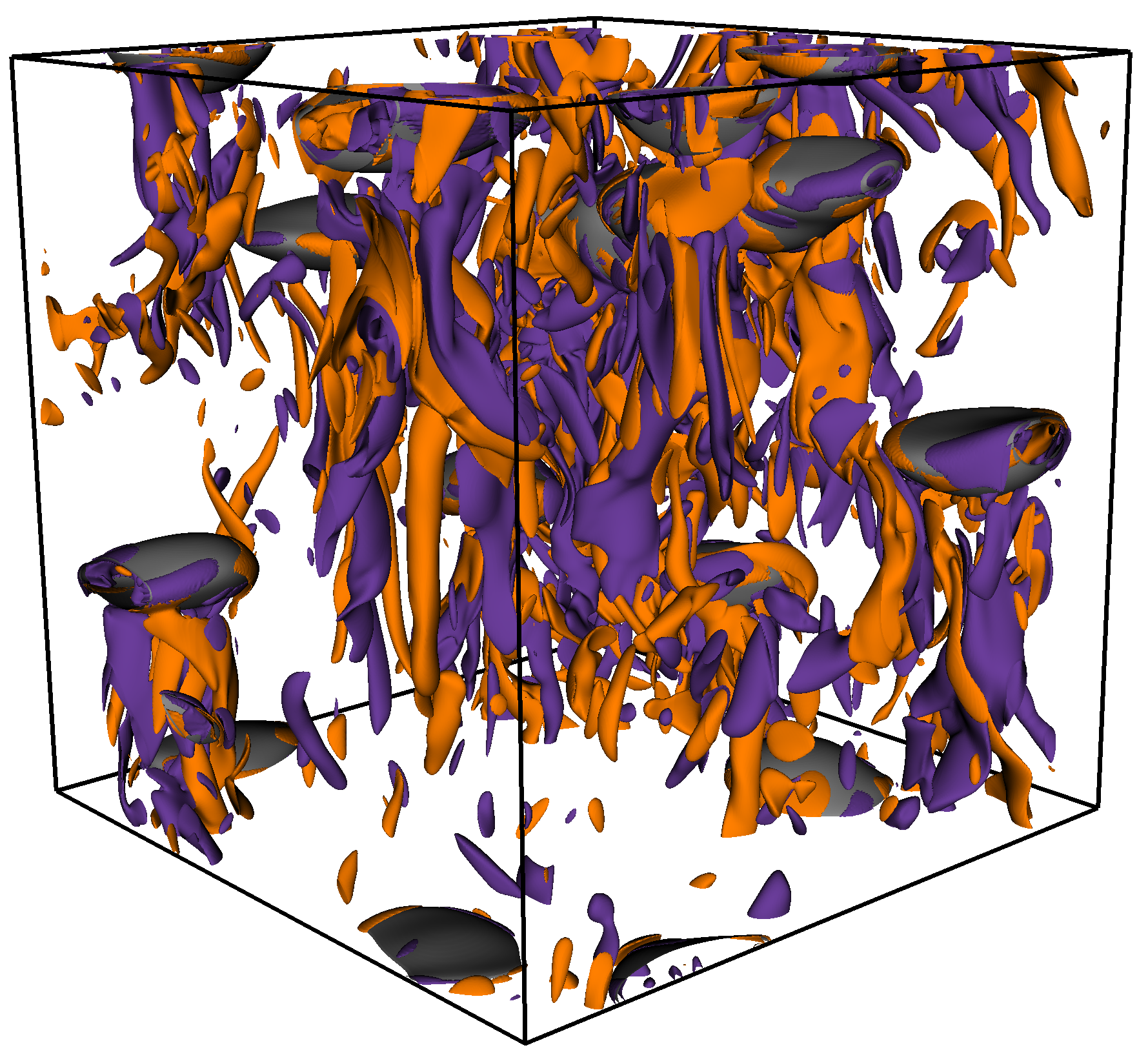}
  \includegraphics[width=0.24\linewidth]{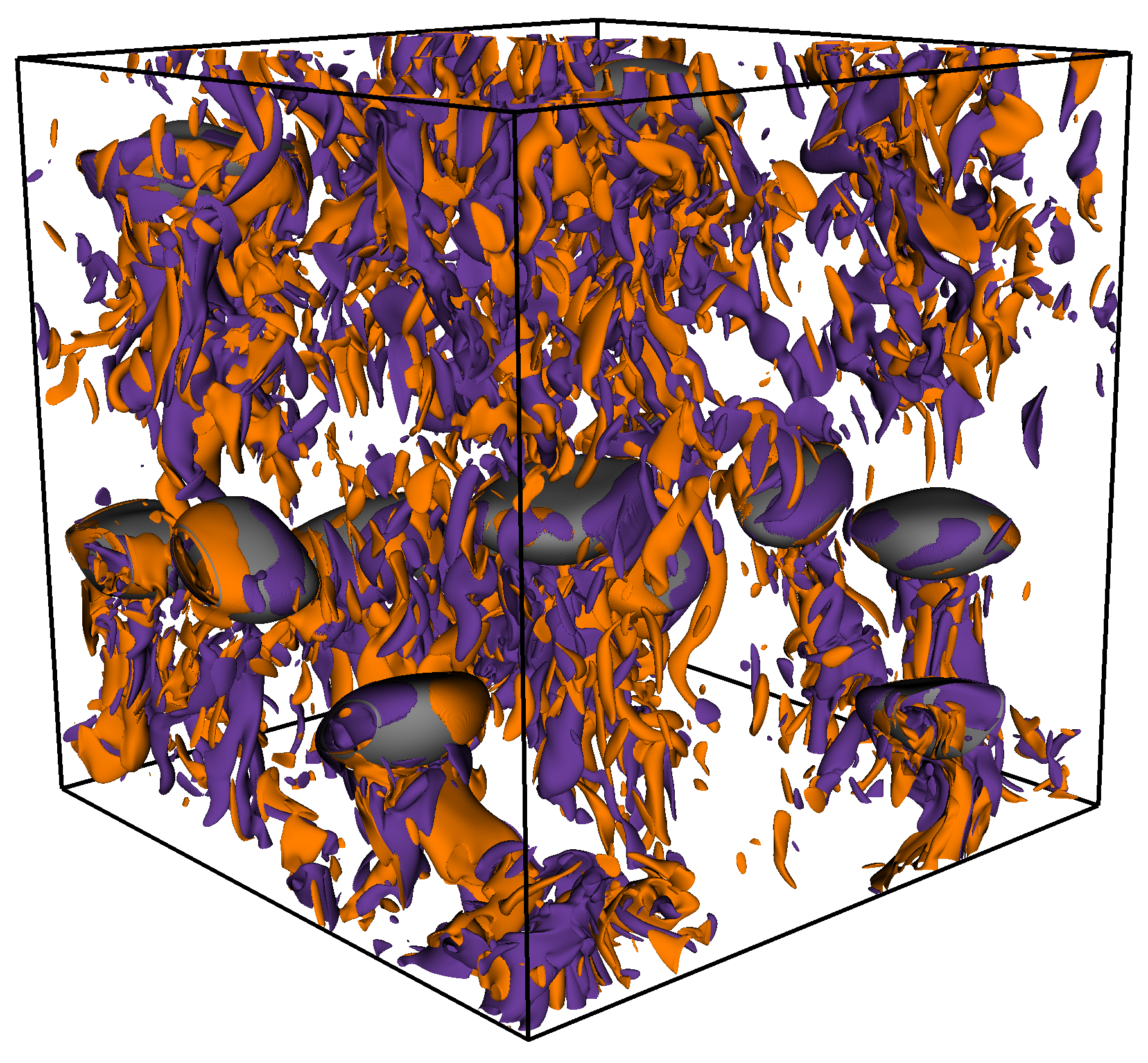}
  \includegraphics[width=0.24\linewidth]{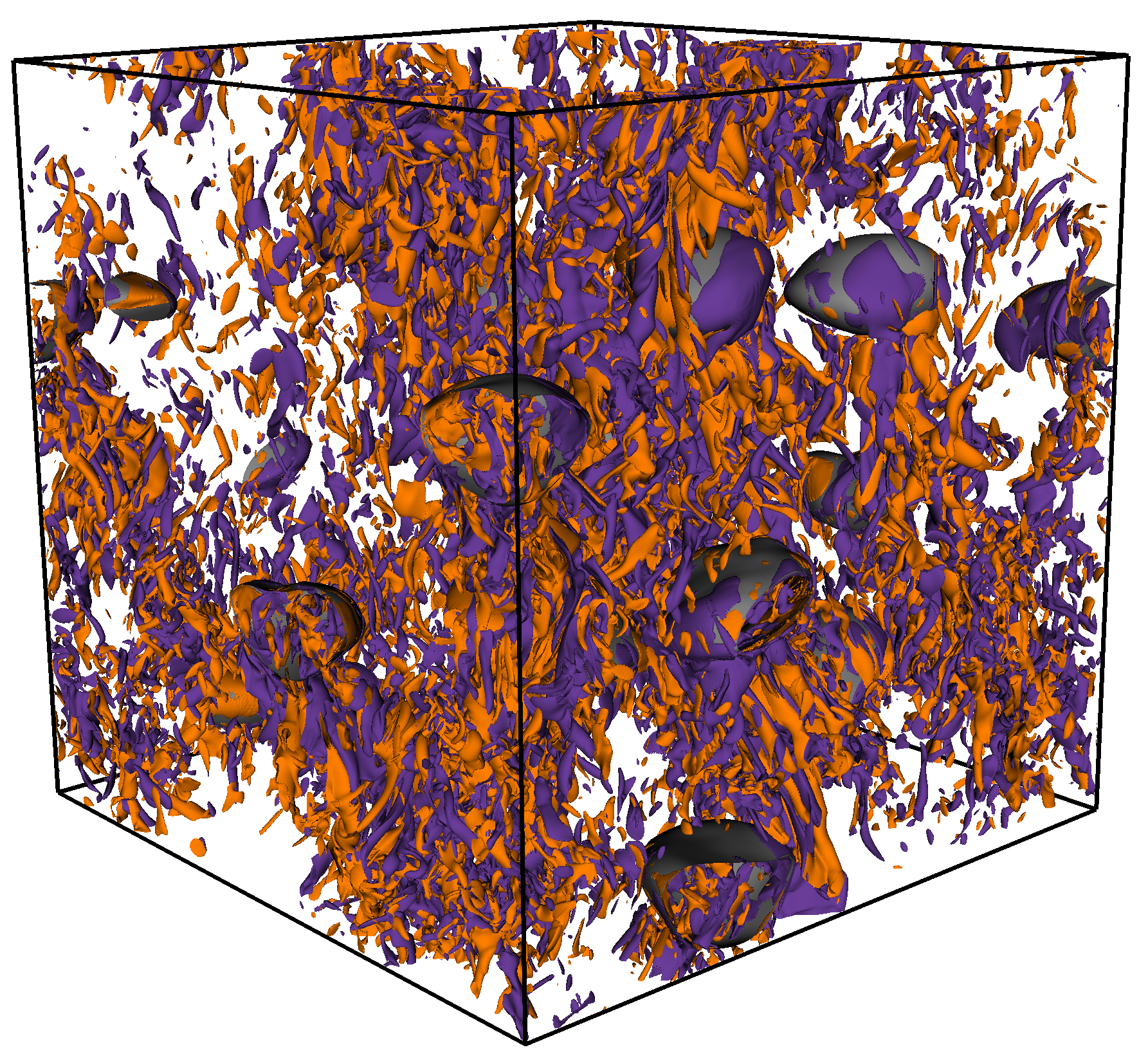}
  \caption{\label{fig:vort}  The iso-contour plot of the z-component of the
      vorticity $\omega^z = \paren{\curl \bu} \cdot \bzh$ for (left to right) $\Ga =
      302,~605,~1029$, and $2057$. We show the contour corresponding to $\pm \braket{
      \paren{\omega^z}^2}^{1/2}$ in purple and orange respectively. The bubbles are
      represented using grey contours.}
\end{figure*}

We study the dynamics of bubbly flow using
multi-phase Navier-Stokes equations~\cite{pandey_2020}
for an incompressible velocity field ${\bu} = (u^x,u^y,u^z)$,
\begin{subequations}
\begin{align}
    \rho(\delt \bu + \bu\cdot\grad\bu) &=  \mu \nabla^2\bu - \grad p +
    \Fst   + \Fg\/,~\text{where} \\
    \Fg \equiv (\rho(c)-\rho_{\rm a})&{\bm g}=\At (\rt+\ro) (c - \ca) {\bm g}\/,~\text{and} \label{eq:Fg}\\
        \Fst \equiv \int \sigma \kappa \bnh \delta(\bx &- \bx_{\rm b})\mathop{\d s}, \label{eq:Fs}
\end{align}
\label{eq:ns}
\end{subequations}
where the bulk viscosity $\mu$ is assumed to be identical in both the phases, and $p$ is
the pressure.
An indicator function $c$, distinguishes the liquid ($c=1$) and the
gas ($c=0$) phase \citep{popinet2018,Tryg2001}.
The density field $\rho(c)=\rt c + \ro(1-c)$. 
In \Eq{eq:ns}, the buoyancy
force is $\Fg$,
$\ca \equiv (1/V)\int c\d V$ is the indicator
function averaged over the volume $V$ of the simulation domain,
$\ra=\ro + (\rt-\ro)\ca$ is the average density, \
${\bg}\equiv -g \bzh$ is the acceleration due to gravity, 
and $\bzh$ is the unit vector along the vertical (positive $z$) direction.
The surface tension force is
denoted by $\Fst$, where  $\sigma$ is the coefficient of the surface tension,
$\kappa$ is the local curvature of the bubble-front located at $\bx_{\rm b}$, $\bnh$
is the unit normal, and $\mathop{\d s}$ is the infinitesimal surface area of the
bubble.

For the small $\At=0.04$ case, we invoke Boussinesq approximation~\cite{chandra1981}
wherein  the density variations can be ignored $\rho \approx \ra$, and the buoyancy force
simplifies to $\Fg=2\ra\At (c-c_a)$ \cite{pandey_2020,pandey_22}.
We solve \Eq{eq:ns} numerically using the pseudo-spectral method~\cite{canuto} in a
three-dimensional periodic domain where each side is of length $L$, 
discretized uniformly into $N$ collocation points.  We numerically integrate the
bubble phase using a front-tracking method \citep{pandey_2020,ramadugu_2020,
Tryg2001}. For time-evolution, we use a second-order exponential time differencing
scheme~\citep{cox_2002} for \Eq{eq:ns} and a second-order Runge-Kutta scheme to
update the front. For the large $\At = 0.8,$ and $0.98$, we use the front tracking module of an open-source
multiphase solver PARIS~\cite{paris}, where both spatial and temporal derivatives are
approximated using a second-order central-difference scheme.

Consistent with the experiments designed to study buoyancy driven bubbly
flows~\cite{vivek_2016,almeras2017,lance_1991} we choose the volume fraction of
the bubbles $\phi \leq 5\%$.
At these volume fraction, the effects coalescence or breakup of the bubbles can be
ignored~\cite{loisy_naso_spelt_2017}.
The front-tracking scheme is ideally suited to study this parameter range because
it ignores both coalescence and breakup.
For one representative case we also perform Volume-of-Fluid (VoF)
simulation ({\tt vof-R5}) using PARIS -- VoF simulations allow coagulation and
break-up -- to confirm that coalescence plays no significant role.

In what follows, the following non-dimensional numbers will be used:
Atwood number $\At$ defined previously,
the Galilei number
$\Ga\equiv\sqrt{(\rho_2 - \rho_1) gd^{3}/\rho_2 \nu^2}$,
the Bond number
$\Bo \equiv {(\rho_2 - \rho_1) g d^2/\sigma}$,
the integral scale Reynolds number
$\ReL^{3/4}\equiv \calL/\eta$,
the Taylor-microscale Reynolds number
$\Rey \equiv \urms^2\sqrt{15  \rho_2/(\nu \epg)}$.
Here we have used,
kinematic viscosity $\nu = \mu/\rho_2$, 
the large eddy turnover time $\tauL \equiv {\cal L}/\urms$,
the root-mean-square velocity, $\urms$, 
the energy injection rate by the buoyant forces \footnote{\ftinjrate},
$\epg\equiv (1/V) \int \Fg \cdot \bu \mathop{\d V}$,
integral length scale 
${\calL}\equiv (3\pi/4) [\sum_k E(k)/k]/\sum_k E(k)$ 
and the Kolmogorov dissipation scale 
$\eta \equiv (\ra \nu^3/\epg)^{1/4}$ 
The different parameters of our simulations are given
in \tabref{tab:runs}.
\begin{table}[!thbp]
    \begin{center}
        \begin{tabular}{lcccccccccccc}
		{\tt runs}  &  \tt{R1}& \tt{R2} & {\tt R3} & \tt{R4} & \tt{R5} & \tt{R6}  &
        {\tt vof-R5} &{\tt R7} & ${\tt R8}$ & ${\tt R9}$\\
            \hline
        $N$ &    $256$ &  $512$ & $512$ & $512$ & $720$ & $720$ & $1024$ & $288$ & $504$ & $504$ \\
        $\Ga$ & $103$  & $302$ & $403$ & $605$ & $1029$ & $2057$ & $1029$ & $340$ & $1059$ & $1489$ \\
        $\At$  & $0.04$  & $0.04$ & $0.04$ & $0.04$ & $0.04$ & $0.04$ & $0.04$ &  $0.8$ & $0.8$ & $0.98$ \\
            $\Rey$ & $11.8$ & $30.7$ & $37.9$ & $47.8$ & $60.7$ & $96.9$ & $60.7$ &
            $28.8$ & $69.3$ & $88.1$ \\
            ${\calL}/\eta$ & $12.6$ & $26.3$ & $29.5$ &  $38.8$ & $50.4$ & $80$ &
            $50.4$ & $20.5$ &  $46$ & $56.0$ \\
        \end{tabular}
        \caption{\label{tab:runs} \textbf{Parameters of simulations.}
          The Bond number $\Bo=1.75$, number of bubbles $\Nb=12$,
          the diameter of the
          bubble $d=1.08$, $L=2\pi$, $\rho_2 =1.0$, and the volume fraction of the bubbles
          $\phi = 3.2\%$ are same in all the runs.
          For all the runs ${\calL} \sim d$, and the statistics are averaged
          over a period of at least $5\tauL$ in  the stationary
          state~(see Supplementary material).
          Total energy injection rate $\epg = 0.031\pm0.002$ in all the case. }

\end{center}
\end{table}

\begin{figure}[!ht]
    \centering
  \includegraphics[width=0.8\linewidth]{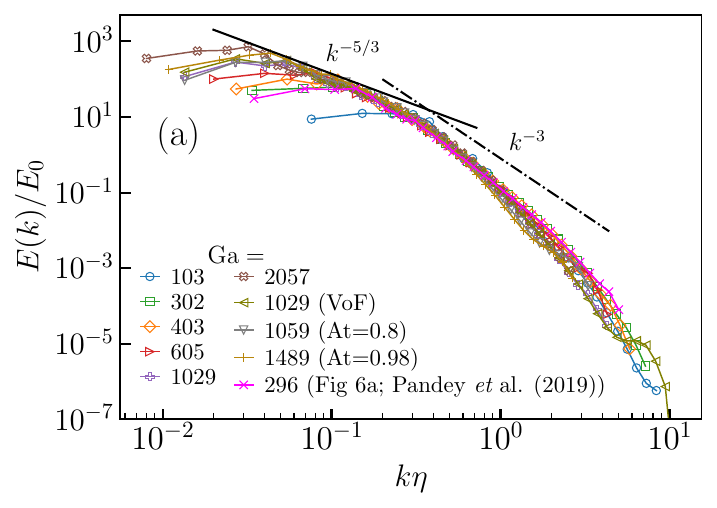}
  \includegraphics[width=0.8\linewidth]{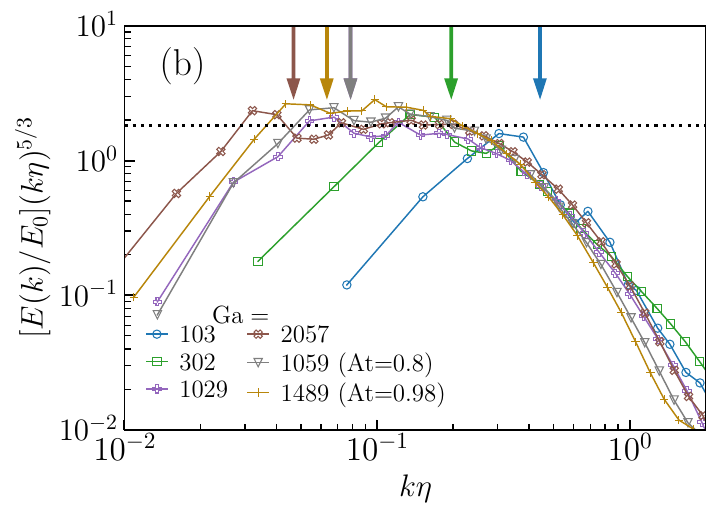}
  \caption{\label{fig:spec1} (a) Log-log plot of the normalized velocity power 
      spectra $E(k)$  as a function of $k\eta$ for various $\Ga$. We observe
      Kolmogorov scaling $E(k)\sim k^{-5/3}$ for  $k\eta \lesssim 0.3$, and the
      pseudo-turbulence scaling $E(k) \sim k^{-3}$ for $k\eta \gtrsim 0.3$.  (b) The
      velocity power spectra for different $\Ga$ compensated by $k^{5/3}$. In (b)
      the vertical arrows show the wavenumber corresponding to the bubble diameter $\kd\eta$. }
\end{figure}
We start our simulation by placing $\Nb$ bubbles randomly in the domain.
  It takes around $4.5\tauL$ for our simulation attain a 
  statistically stationary state.
Once it is reached, all our data are averaged over at least $5\tauL$. 
In \fig{fig:vort} we show the iso-contour plot of the $z$-component of vorticity
$\omega^z = \paren{\curl \bu }\cdot \bzh$. As the $\Ga$ ($\Rey$) is increased, not
only the intense vortical regions in the field increases, we observe flow structures
at much smaller scales as well.  

Next we investigate power spectrum of velocity fluctuations:
\begin{equation}
E(k)=\frac{1}{2}\sum_{\bk}\braket{\buh(\bk)\buh(-\bk)}\delta(|\bk| - k), 
\label{eq:spec}
\end{equation}
where $\buh(\bk)$ is the Fourier transform of the velocity field $\bu$,
$\bk$ the wavevector
and $\braket{\cdot}$ denotes spatiotemporal average over the statistically stationary
state of turbulence. 
Kolmogorov theory of turbulence shows that,
in homogeneous and isotropic turbulence, 
$E(k)$ for different Reynolds numbers collapses
onto a single curve if we use $\eta$ 
as the characteristic length scale and
$\Enot = \paren{\epg\nu^5/\rho_2}^{1/4}$ as the characteristic
energy scale, which we use henceforth. 
In \subfig{fig:spec1}{a} we show that, even for buoyancy
driven bubbly  turbulence,  the same data--collapse holds
for scales $k \gtrsim \kd$.
For small $\Ga$ number we obtain the pseudo--turbulence
regime~\cite{lance_1991,vivek_2016,almeras2017,pandey_2020}
for $k \gtrsim 0.3/\eta$. 
As the $\Ga$ increases an scaling range with an exponent of
approximately $-5/3$ emerges
for $\kd \lesssim k  \lesssim 0.3/\eta$.
This is  a novel, previously unobserved scaling in bubbly flows.
The scaling  range increases with  $\Ga$;
it is almost non-existent for $\Ga=100$ and extends
up to almost half a decade for $\Ga=2057$.  
The $-5/3$ scaling range is best seen in \subfig{fig:spec1}{b} where we plot the
spectra compensated with $k^{5/3}$.
As we have used $\eta$ as our characteristic length scale the Fourier mode $\kd$,
shown by an arrow appears at different locations in this plot.
As $\Ga$ is increased $\kd$ moves to the left thereby the
$-5/3$ scaling range emerges.

Note that due to rising bubbles, in principle, our flow is anisotropic.
Here and henceforth, following the standard practice in bubbly 
turbulence \cite{lance_1991,pandey_2020,chibbaro_2021},
we use the isotropic spectra which is the projection
of the general anisotropic spectra on to the isotropic
sector~\cite{biferale2005anisotropy}. 
In the supplementary material, {which includes
Ref.~\cite{biferale2005anisotropy}}, we show that for our simulations the
anisotropic contribution is negligible at all scales except $k$ in the
neighbourhood of $\kd$.

We now describe how Kolmogorov scaling emerges
at both small and large $\At$ by studying the scale-by-scale energy
budget equation:
\begin{equation}
  \delt \mEK = -\PiK - \FsK + \PK  -\DK + \FgK\/.
  \label{eq:ebud}
\end{equation}
Here $\mEK$ is the kinetic energy contained up-to wavenumber $K$.
Here $\PiK$, $\FsK$, $\PK$, $\DK$ and $\FgK$ are the contributions from the
advective term, surface tension, pressure, viscous dissipation and
buoyancy from \Eq{eq:ns}~\footnote{See Supplementary Material for detailed 
step-by-step derivation, which includes Refs.~\cite{pope,eyink_1995,frisch,aluie_2013,aluie_2019}}.

\begin{figure}[!t]
    \centering
    \includegraphics[width=1.01\linewidth]{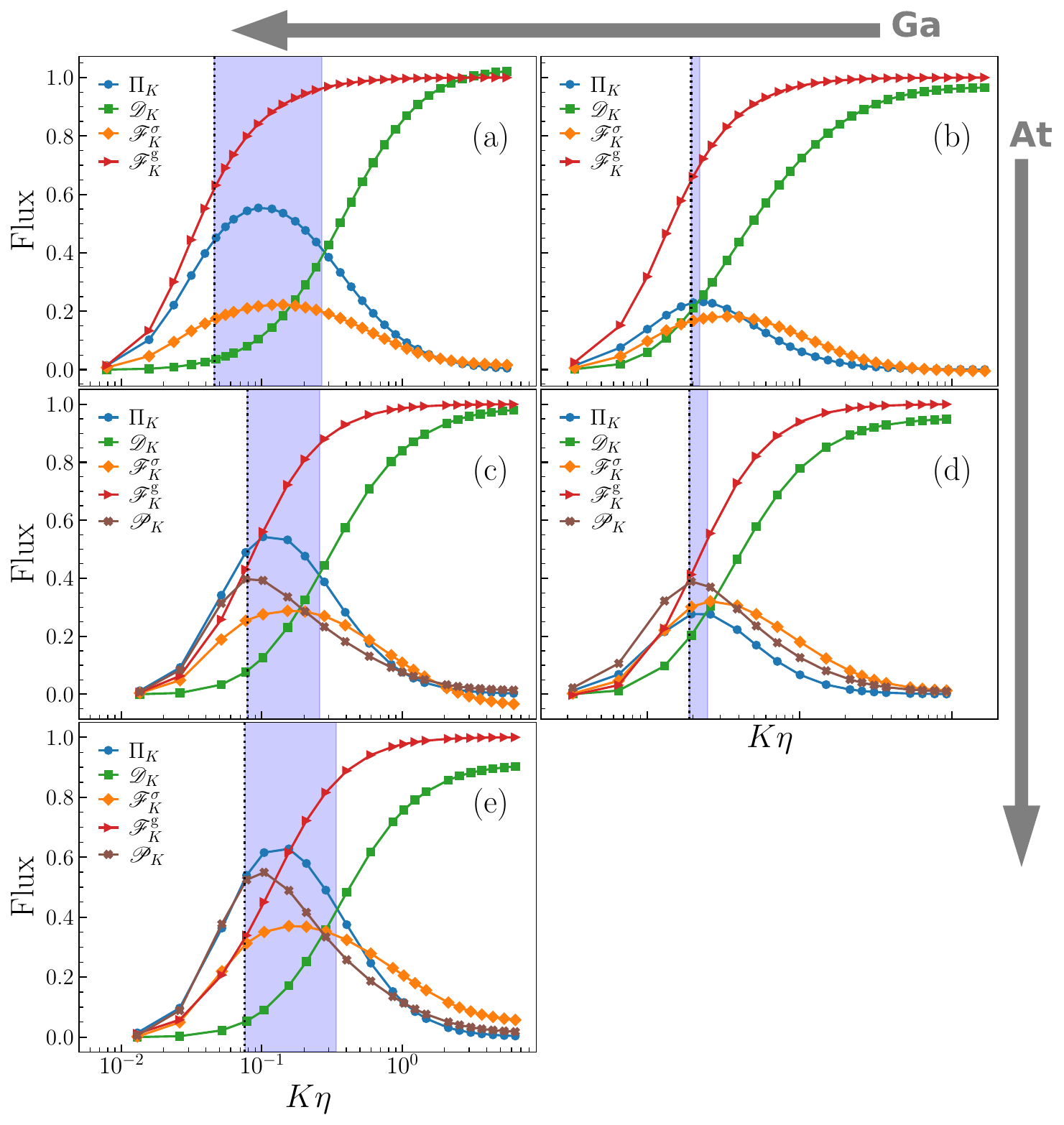}
    \caption{\label{fig:flux} Semi-log plot of the different terms in
      scale-by-scale budget  for low Atwood number $\At=0.04$ (top row)
      with $\Ga=2057$ (a) and $\Ga=302$ (b);
       $\At=0.8$ (middle row) for $\Ga=1029$ (c), $\Ga=345$ (d); and 
       $\At=0.98$ (bottom row), $\Ga=1489$ (e).
      The abscissa is normalized by the  total energy injection rate $\epg$.
      The dashed vertical line represents the wave-number corresponding to bubble
      diameter, $\kd$. The continuous vertical line shows the wavenumber 
      where the advective flux and the dissipative flux crosses.
      The region between these two lines, where the advective
      contribution dominates, is shaded with light blue color.}
\end{figure}

{\it The scale-by-scale budget for low ${\At=0.04}$} ---
we follow Refs.~\cite{pope,eyink_1995,frisch,verma_2019} to derive \Eq{eq:ebud}.
We consider stationary state, hence $\delt \mEK=0 $ and
we use Boussinesq approximation, hence $\PK = 0$.
We plot all the others terms of \Eq{eq:ebud} as a function of $K$
in the top row of \fig{fig:flux} for large and small $\Ga$. 
As expected~\cite{pandey_2020}, bubbles 
inject energy into the flow at scales comparable to the bubble diameter
--  $\FgK$ monotonically increases and saturates around $K\approx \kd$.
From the perspective of the  Kolmogorov theory
of turbulence~\cite[][section 6.2.4]{frisch} the buoyancy injection term $\FgK$ is
the large scale driving force active at scales around $\kd$. Following
Ref.~\cite{frisch}, consider a fixed $K \gg \kd$ and take the limit $\nu\to 0$
($\Ga\to\infty$).  Then $ \displaystyle \lim_{\nu \to 0}\DK \approx 0$, holds and
the flux balance equation reads:
\begin{equation}
  \PiK + \FsK  = \epg\/.
  \label{eq:inner_range}
  \end{equation}
Because the injection is limited to Fourier modes around $\kd$, for $K \gg \kd$,
$\FgK \approx \epg$ is a constant.
In homogeneous and isotropic turbulence in absence of bubbles the dissipative effects 
become significant around
$8$ to $10\eta$~\cite{yaglom}.  
We find $3\eta$ is a reasonable approximation in our case.
Thus, \Eq{eq:inner_range} is expected to be valid for $\kd < K \lesssim 0.3/\eta$
-- this range is shaded with light blue in \fig{fig:flux}.
Within the shaded region $\PiK \gg \FsK$, hence
$\PiK\approx \epg/2$ is a constant leading to the
Kolmogorov $-5/3$  scaling in the energy spectrum~\cite{frisch}.
Even at $\Ga=2057$, the $-5/3$ scaling range is at best close
to a decade.
In \subfig{fig:flux}{b}, for $\Ga = 302$ the shaded region
has practically disappeared.
For this and other other runs with  smaller $\Ga$, we expect to observe
pseudo-turbulence where none of the three fluxes, $\FsK$, $\PiK$ and $\DK$, can be
ignored.  A detailed discussion on
the flux balance in the pseudo-turbulence regime for $\Ga \leq 360$ can be
found in our earlier studies~\cite{pandey_22,pandey_2020,ramadugu_2020}.

{\it The scale-by-scale budget for high ${\At=0.8,0.98}$} --- 
  we follow Refs.~\cite{pope,aluie_2013,aluie_2019} to derive \Eq{eq:ebud}.
  We again consider statistical stationarity, hence $\delt \mEK=0 $.
  In \subfig{fig:flux}{c-e}, we
  plot all the terms of \Eq{eq:ebud} as a function of $K$
  for both high and low $\Ga$.
  The  ``baropycnal work", $\PK$, now provide an alternate routes for
  nonlinear energy transfer. The baropycnal term has contributions from the
  barotropic generation of strain and baroclinic generation of vorticity due to
  density variations \cite{aluie_2019}.
  Remarkably, for large enough $\Ga$ there is a range of scales, shaded in
  \subfig{fig:flux}{c,e} where the dominant balance is
  $\PiK \approx \epg/2$ is a constant leading to the Kolmogorov
  $-5/3$  scaling in the energy spectrum.
  The other transfer mechanisms $\FsK$ and $\PK$ are sub-dominant.
  A positive slope of $\FsK$ ($\PK$)  indicates that the energy is
  absorbed (injected), whereas a negative slope indicates energy being
  injected (absorbed).
  Thus surface tension absorbs energy at large scales and injects it at small
  scales, whereas the opposite is the case for the baropycnal term.

Several earlier experimental and numerical
studies~\cite{risso_legendre_2010,martinez_2010,roghair,vivek_2016,
  pandey_2020,chibbaro_2021,Ma_2021,Ma_2022,pandey_22} have
shown that the power spectrum of velocity fluctuations is insensitive to
variation in $\At$ for $\Ga \lessapprox 350$.
We have now shown that this is also true for large $\Ga$.
Thus, the following scenario emerges. For a fixed but small $\At$, where
the Boussinesq approximation is valid, the baropycnal flux is negligible.
For a fixed but large $\At$ it is not.
But even for the latter case  as the Galilei number $\Ga$ is increased
beyond some critical value $\Ga^{\ast}$,
the advective flux can become the dominant contribution
to the net flux. In such cases, Kolmogorov-like scaling holds.
A systematic study to find out the how $\Ga^{\ast}$
depends on $\At$ is outside the scope of this work.

We comment that the resolution required to conduct a fully  resolved
    pseudo-turbulent simulation increases proportionally with both $\At$ and 
    $\Ga$ \cite{cano2016,chibbaro_2021}.  However, a comparison of different
    experimental and numerical
    studies~\cite{pandey_2020,chibbaro_2021,Ma_2021,Ma_2022,pandey_22} reveals
    that the statistics of the velocity fluctuations, in particular the PDF and
     the power spectra, are robust to the variation in both $\At$ and grid resolution.
    The effect of resolution is only observed at very small scales ({see supplementary 
    material, which includes Ref.~\cite{chibbaro_2021}}) and therefore we expect all our 
    results will be valid at resolutions higher than the current study.

To conclude, we  demonstrate, for the first time, that at large enough
$\Ga>1000$, the power spectrum of velocity fluctuations shows the Kolmogorov
scaling for range of scales  between the bubble diameter and the dissipation scale.
For scales smaller than  $\eta$, the physics of pseudo-turbulence is recovered.
Most of the earlier experiments  on buoyancy driven bubbly flows have considered
air bubbles of diameter $d \lesssim 5$mm in water, which correspond to
$\Ga \lesssim 1000$~\cite{vivek_2016,risso_legendre_2010}.
Our study suggests that experiments
with air bubbles of diameter $d \geq 7.5$mm are needed to achieve $\Ga>1000$ and 
observe the Kolmogorov scaling range.  
At both high and low Atwood, we expect the $-5/3$ scaling range to increase
further as the $\Ga$ is increased. Due to the various computational
challenges \cite{chibbaro_2021}, although such a study is currently not
possible, it demands future investigations.

\acknowledgements
{DM and VP acknowledge the support of the Swedish Research Council Grant
    No.
    638-2013-9243 and 2016-05225.  Nordita is partially supported by Nordforsk.
    PP and VP  acknowledge support from  the Department of Atomic Energy (DAE),
    India under Project Identification No. RTI 4007, and DST (India) Project
    Nos. MTR/2022/000867, DST/NSM/R\&D\_HPC\_Applications/2021/29, and
    DST/NSM/R\&D\_HPC\_Applications/Extension/2023/08. Most of the simulations
    are done using the HPC facility at TIFR Hyderabad, and the National
    Supercomputing Mission facility (Param Shakti) at IIT Kharagpur.  Some of
    the simulations were performed on resources provided by the Swedish
    National Infrastructure for Computing (SNIC) at PDC center for high
performance computing.}


\newpage
\clearpage
\section*{Supplemental Material}
\section{Characterizing anisotropy  in buoyancy driven bubbly flows}
 In the following section we characterize the anisotropy in buoyancy driven bubbly flows. We 
 show that the anisotropic contribution to the velocity correlations are dominant at scales larger 
 than the bubble diameter. Thus we are justified the use of spherical averaged energy spectrum in the 
 to study bubbly flows.

\subsection{Liquid velocity fluctuations}
\begin{figure}[!h]
  \includegraphics[width=1.02\linewidth]{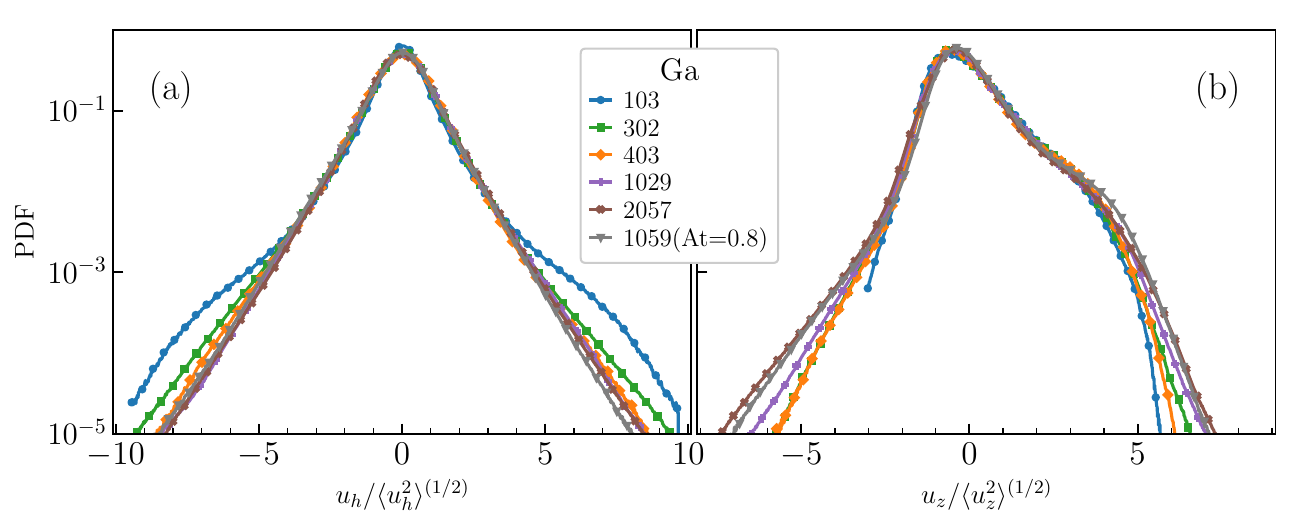}
  \caption{\label{fig:hist} The probability distribution function of the (a) horizontal
       and (b) the vertical component of the liquid velocity fluctuations at
   different $\Ga$. }
\end{figure}

In \fig{fig:hist} we plot the probability distribution function (PDF) of the liquid
velocity fluctuations for different $\Ga$. The PDF of the
vertical ($u^z$) velocity fluctuations  are skewed as we expect more positive
fluctuation in the  wake of the bubbles~\cite{vivek_2016,pandey_2020}. The PDF of
horizontal components is symmetric~\cite{riss18,chibbaro_2021,pandey_2020}. These
PDF are consistent with what has been observed in earlier experiments and
simulations at smaller $\Ga$~\cite{riss18}.  We remark that, even though the flow
fields at various $\Ga$ are visually different from one another, the shape of the
distribution remains the same.

\subsection{Velocity power spectra}
The two-point velocity correlations can be characterized in Fourier space using
the second rank spectral tensor 
\begin{align}
C^{\alpha\beta}(\bk) \equiv \braket{\uh^{\alpha}(\bk)\uh^{\beta}(-\bk)},
\end{align}
where indices $\alpha,\beta=x,y,z$. The power spectrum of velocity fluctuations can 
be rewritten in terms of the spectral tensor as
\begin{align}
E(k) = \frac{1}{2}\sum_{\bk} C^{\alpha \alpha}({\kk}) \delta(|\bk| - k).
\end{align}

As the anisotropy in bubbly flows is because the buoyancy force is along the
$\zh$-direction. Therefore, we use  the axisymmetric turbulence formalism
outlined in \cite{biferale2005anisotropy} and construct two unit vectors
orthogonal to ${\bk}$:
  \begin{align}
      \eeo \equiv \frac{\bk\times\zh}{\lvert \bk\times\zh \rvert} \/, \quad {\rm and}~ 
    \eet \equiv \frac{\bk\times(\bk\times\zh)}{\lvert \bk\times(\kk\times\zh) \rvert} \/.
\label{eq:vec}
\end{align}

The spectral tensor can be written in terms of the $\eeo$ and $\eet$ vectors as
\begin{align}
C^{\alpha\beta}(\kk)  = A(\kk)\eo^{\alpha}\eo^{\beta} + B(\kk) \et^{\alpha}\et^{\beta},
\label{eq:cab}
\end{align}
with indices $\alpha,\beta=x,y,z$. 

We evaluate $C^{\alpha \beta}(\bk)$ from our DNS, and use \Eq{eq:vec}, and (\ref{eq:cab}) to 
obtain
\begin{align}
\nonumber
A(\kk) &=  \frac{k_y^2 C^{xx}(\kk) - 2 k_x k_y \operatorname{Re}[C^{xy}(\kk)] + k_x^2 C^{yy}(\kk)}{k_x^2+k_y^2},\quad ~\rm{and} \/ \\
\nonumber
B(\kk) &= C^{zz}(\kk) \frac{k^2}{k_x^2 + k_y^2}.
\nonumber
\end{align}

Note that the function $A(\kk)$ gets contribution only from the horizontal
velocity fluctuations, whereas $B(\kk)$ only depends on vertical velocity
fluctuations. By performing the angular averaging, similar to \Eq{eq:spec}~(main document)
we define the one-dimensional spectra
\begin{align}
a(k) &= \frac{1}{2}\sum_{\bk} A({\kk}) \delta(|\bk| - k)~{\rm and~}\\
b(k) &= \frac{1}{2}\sum_{\bk} B({\kk}) \delta(|\bk| - k).
\end{align}

\begin{figure}[htpb]
    \centering
    \includegraphics[width=0.48\textwidth]{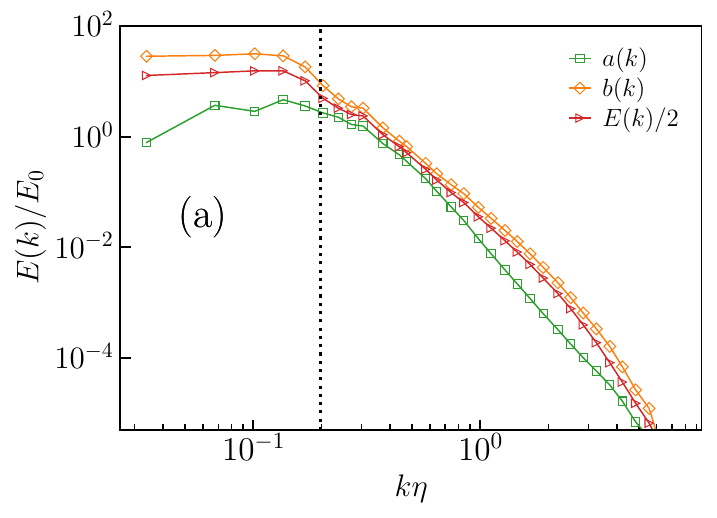}
    \includegraphics[width=0.48\textwidth]{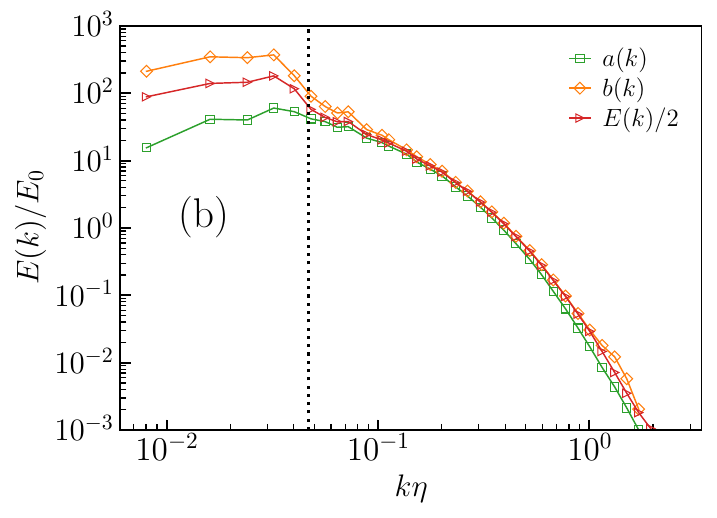}
    \caption{Anisotropic decomposition of velocity power spectra for (a) $\Ga=302$,
    (b) $\Ga=2057$. Vertical line show the wave mode corresponding to the bubble diameter.}
    \label{fig:aniso}
\end{figure}

We expect $a(k)=b(k)$ for homogeneous, isotropic turbulence.  In
\fig{fig:aniso} we compare different spectrum  for $\Ga=302$ and $2057$. The
flow isotropy is higher at scales larger than the bubble diameter.  For small
$\Ga=302$, most of the contribution to the energy spectrum comes from the
vertical velocity fluctuations ($E(k)\approx 2b(k)>a(k)$). However, all the
spectrum show identical scaling behaviour. On increasing the $\Ga=2057$, we
find that $E(k)\approx 2a(k) \approx 2b(k)$ for scales larger than the bubble
diameter indicating isotropization of small scale fluctuations. Therefore, we
conclude that \Eq{eq:spec}~(main document) is a good indicator to study the scaling
behaviour of velocity fluctuations in bubbly flows.

 \section{Derivation of the scale-by-scale budget equations}
In this section, we detail the complete derivation of the scale-by-scale energy budget equations.
Following Ref.~\cite{pope,eyink_1995,frisch,aluie_2013}, for any field, $\psi(\bx)$  we obtain 
the corresponding field filtered at scale $K$ as,
\begin{subequations}
\begin{align}
    \fil{\psi}(\bx) &\equiv \int\exp(i\bq\cdot\bx) \GK(\bq)\hat{\psi}({\bq}) {\mathop{\d\bq}}\/,
    \quad\text{where,}
  \label{eq:filop} \\
    \GK(\bq) &\equiv \exp\paren{-{\pi^2q^2\over 24 K^2}}\/   \label{eq:kernal}
    \end{align}
\end{subequations}
where $\hat{\psi}(\bq)$ is the Fourier transforms of $\psi$, and $\GK$ is a low-pass filtering 
kernel which is smooth in both physical and Fourier space.
As we are dealing with a flow with density fluctuations, we define Favre filtered field 
${\til{\psi}} = \fil{(\rho \psi)}/\fil{\rho}$ \cite{aluie_2013,aluie_2019}.

We define the cumulative energy flux through scale $K$ as
$\mE_K = {1\over2}\braket{\fil{\rho}\mid\til{\bu}\mid^2}$,
such that the spectrum $E(k) \sim (1/\rho_2) \partial_K \mE_K\mid_{K=k}$.
In the statistically stationary state $\partial_t \mE_K = 0$. Thus,
from the Navier-Stokes equation Eqs.~\eqref{eq:ns}~(main document), and following the procedure outlined
in Refs.~\cite{pope,aluie_2019}  we obtain the following scale-by-scale kinetic 
energy budget equation in the statistically stationary state:

\begin{subequations}
    \begin{align}
        \PiK  &+ \FsK - \PK = -\DK + \FgK\/,\text{where} \label{eq:bud1}\\
        \PiK &\equiv  -\braket{\fil{\rho}{\delb \til{u}^\alpha} \tau^{\ab}_K}\/, \\
        \FsK &\equiv -\braket{\til{\bu}\cdot{\fil{\bm{F}}^\sigma}}\/,\\
        \PK &\equiv - \braket{\til{\uu} \cdot {\fil{(\nabla p)}}} \/,\\
        \DK &\equiv 2 \mu \braket{\delb \til{u}^\alpha \fil{S}^{\ab}}\/, \\
        \FgK &\equiv \braket{\til{\bu}\cdot\fil{\bm{F}}^g} \/, \text{where}\\
        S^{\ab} &\equiv \frac{1}{2}\left[ \dela\ub + \delb \ua)\right]\/, \text{and} \\
        \tau^{\ab}_K &\equiv \til{(u^{\alpha}u^{\beta})} - \til{u}^{\alpha} \til{u}^{\beta}.
    \end{align}
   \label{eq:bud1}
\end{subequations}

In \Eq{eq:bud1}, $\PiK$ is the advective flux, $\tau_K$
is the Reynolds stress tensor. In bubbly flows, the  
``baropycnal work" $\PK$ and the surface tension term $\FsK $ 
provide alternate routes for nonlinear energy
transfers. The baropycnal term has contributions from the barotropic 
generation of strain and baroclinic generation of vorticity due to
density variations \cite{aluie_2019}.  The other terms in \Eq{eq:bud1} are 
the cumulative injection rate  up to
wavenumber $K$ due to buoyancy $\FgK$, and dissipation rate up to wavenumber $K$, $\DK$.

At low Atwood number, we can employ Boussinesq approximation. Therefore
$\til{{\bm u}}\approx\fil{\bm u}$, similarly  $\mE_K \approx {1\over2}\ra
\braket{\mid\fil{\bu}\mid^2}$, and the power spectrum $E(k) = (1/\ra)\partial_K
\mE_K\mid_{K=k}$. The other terms in the budget equation reduces to:

\begin{subequations}
    \begin{align}
        \PiK &\equiv  -\ra \braket{ \fil{S}^\ab \tau^{\ab}_K}\/, \\
        \FsK &\equiv -\braket{\fil{\bu}\cdot{\fil{\bm{F}}^\sigma}}\/,\\
        \PK &=0\/, \\
        \DK &\equiv 2 \mu \braket{\fil{S}^{\ab} \fil{S}^{\ab}}\/, \\
        \FgK &\equiv \braket{\fil{\bu}\cdot\fil{\bm{F}}^g} \/, \text{and}\\
        \tau^{\ab}_K &\equiv \fil{(u^{\alpha}u^{\beta})} - \fil{u}^{\alpha} \fil{u}^{\beta}.
    \end{align}
   \label{eq:budlat}
\end{subequations}

\section{Statistically stationary state}

\begin{figure}[!h]
  \includegraphics[width=0.6\linewidth]{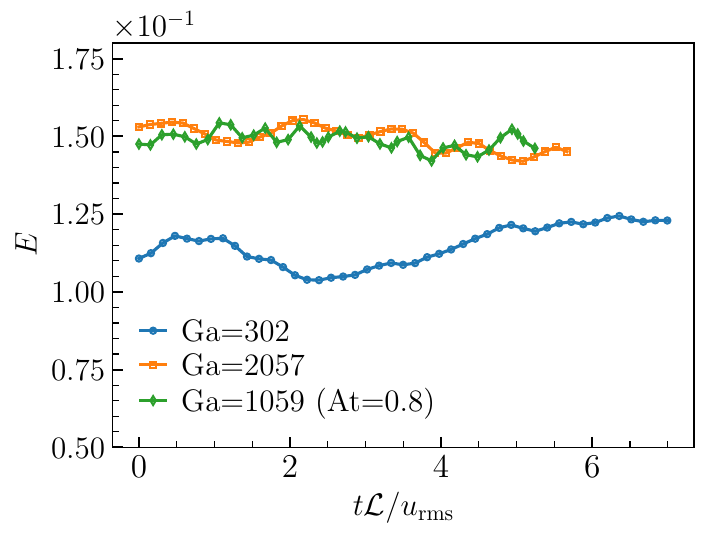}
  \caption{\label{fig:stationary} The time evolution of $E(t)$ for $\Ga=302, 2057$ at 
  $\At=0.04$ and $\Ga=1059$ at $\At=0.80$.}%
\end{figure}

In \fig{fig:stationary}, we show the time series of $E(t) = \overline{\bu^2}/2$ 
for a time period over which we have averaged the data for a few representative 
simulations. Note that within the scope of the current section, 
$\overline {(\cdot)}$  represents spatial average.  

\section{Resolution test}
\begin{figure}[!h]
    \centering
    \includegraphics[width=0.48\linewidth]{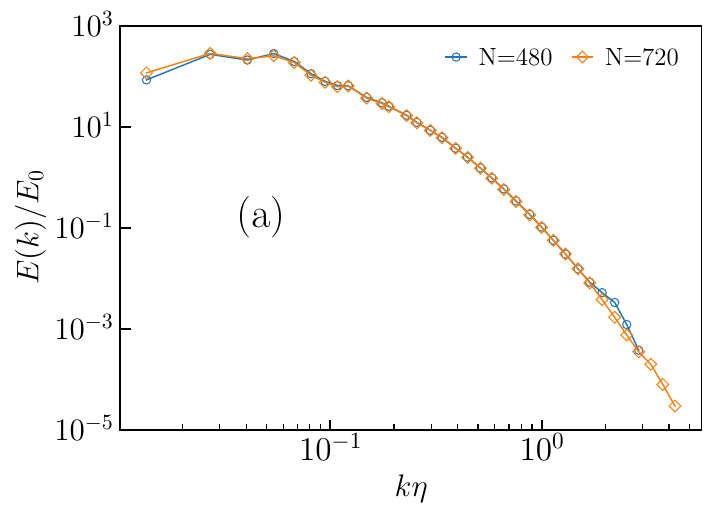}
    \includegraphics[width=0.48\linewidth]{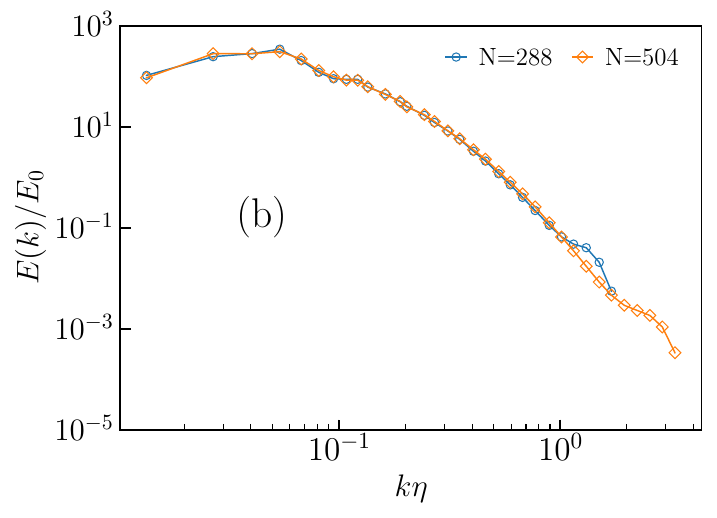}
    \caption{The comparison of velocity power spectrum for (a) $\At=0.04, \Ga=1029$ at spatial
        grid resolution of $N=480{ ~\rm and}, 720$, (b) $\At=0.8, \Ga=1059$ at spatial
        resolution of $N=288{ ~\rm and}, 504$.}%
    \label{fig:resln}
\end{figure}

We show comparison of power spectra at $\At=0.04$ 
$\Ga=1029$ at resolution $N=480$ and $720$ in \fig{fig:resln}.
Similarly we also show the spectra for high $\At=0.8$ 
$\Ga=1059$ at resolution $N=288$ and $504$. We find that in both the cases 
scaling ranges to be well resolved as the spectra at different resolution overlaps.
We observe a kink at the tail of the spectra indicating very small scales at deep dissipation range
are under-resolved. As the resolution is increased this kink gets pushed to at even smaller scales, 
extending the pseudo-turbulent scaling. The resolution required to resolve these scales in the deep dissipation
range increases with $\Ga$ \cite{chibbaro_2021} and is beyond the scope of current work.

\bibliographystyle{apsrev4-1.bst}
\bibliography{hga_paper.bib}
\end{document}